# Large-scale jets in the magnetosheath and plasma penetration across the magnetopause: THEMIS observations

A. V. Dmitriev[1,2] and A. V. Suvorova[1,2]

[1]*Institute of Space Sciences, National Central University, Chung Li, Taiwan*

[2]*Skobeltsyn Institute of Nuclear Physics, Moscow State University, Moscow, Russia*

Short title: PENETRATION OF LARGE-SCALE JETS

---

A. V. Dmitriev, Institute of Space Science National Central University, Chung-Li, 320, Taiwan, also D.V. Skobeltsyn Institute of Nuclear Physics, Moscow State University, Russia (e-mail: dalex@jupiter.ss.ncu.edu.tw)

A. V. Suvorova, Institute of Space Science National Central University, Chung-Li, 320, Taiwan, also D.V. Skobeltsyn Institute of Nuclear Physics, Moscow State University, Russian Federation (e-mail: suvorova_alla@yahoo.com)

**Abstract** THEMIS multi-point observation of the plasma and magnetic fields, conducted simultaneously in the dayside magnetosheath and magnetosphere, were used to collect 646 large-scale magnetosheath plasma jets interacting with the magnetopause. The jets were identified as dense and fast streams of the magnetosheath plasma whose energy density is higher than that of the upstream solar wind. The jet interaction with the magnetopause was revealed from sudden inward motion of the magnetopause and an enhancement in the geomagnetic field. The penetration was determined as appearance of the magnetosheath plasma against the background of the hot magnetospheric particle population. We found that almost 60% of the jets penetrated through the magnetopause. Vast majority of the penetrating jets was characterized by high velocities $V > 220$ km/s and kinetic $\beta_k > 1$ that corresponded to a combination of finite Larmor radius effect with a mechanisms of impulsive penetration. The average plasma flux in the penetrating jets was found to be 1.5 times larger than the average plasma flux of the solar wind. The average rate of jet-related penetration of the magnetosheath plasma into the dayside magnetosphere was estimated to be ~$10^{29}$ particles per day. The rate varies highly with time and can achieve values of $1.5*10^{29}$ particles per hour that is comparable with estimates of the total amount of plasma entering the dayside magnetosphere.

*Keywords:* magnetosheath, plasma jets, magnetopause, plasma transport into the magnetosphere

# 1. Introduction

The terrestrial magnetopause is a current layer separating the Earth's magnetosphere from the solar wind plasma streams. Before interaction with the magnetopause, the solar wind streams of high kinetic energy are slowed down and become turbulent in the magnetosheath such that the solar wind kinetic energy is converted into the thermal and magnetic energy of the magnetosheath plasma [*e.g. Song et al.,* 1999]. As a result, the magnetopause is affected by dense plasma of 0.1 to 1 keV thermal energy and highly variable interplanetary magnetic field (IMF) amplified in the magnetosheath. At low latitudes on the dayside, the average magnetopause thickness is about 500 km [*Berchem and Russell*, 1982] that corresponds to tens of gyro radii for the protons with energy of several hundreds of eV. Hence, the magnetopause should be impenetrable for the magnetosheath plasma.

However under certain circumstances, the magnetosheath plasma can penetrate across the low-latitude dayside magnetopause. One of most known processes is magnetic reconnection [e.g. *Sibeck*, 1999; *Eriksson et al.*, 2009; *Hasegawa*, 2012]. Another phenomenon is high-$\beta$ magnetosheath plasmoids or jets [*Lemaire*, 1985; *Savin et al.*, 2008; *Dmitriev and Suvorova*, 2012]. Magnetosheath plasma jets are defined as intense localized fast ion fluxes whose kinetic energy density can be several times higher than that in the upstream solar wind. In recent years, numerous authors reported observations of jets by Cluster and THEMIS missions [*Dmitriev and Suvorova*, 2012 and references therein; *Karlsson et al.*, 2012; *Plaschke et al.*, 2013; *Gunell et al.*, 2012; 2014].

A number of mechanisms have been proposed for penetration of the magnetosheath plasmoids and plasma jets through the magnetopause [see *Sibeck*, 1999 and references therein]. One of them is so-called finite Larmor radius (FLR) effect [e.g. *Savin et al.*, 2008]. On the base of Cluster observations, it was suggested that jets can pierce through the high-latitude magnetopause due to a kinetic effect for the penetration of magnetosheath ions with an energy of >350 eV through a thin current sheet of ~90 km thickness. However, the FLR mechanism was hard to apply for the low-latitude dayside magnetopause of ~500 km thickness. Another mechanism is an impulsive penetration [*Lemaire*, 1977;

1985; *Echim and Lemaire*, 2000]. The mechanism is based on an idea that the upstream solar wind is non-stationary and not uniform that leads to the formation of localized pressure pulses along the bow shock. The pressure pulses propagate across the magnetosheath in the form of plasma irregularities or plasmoids, which can impact the magnetopause. Plasmoids with an excess momentum density penetrate into the magnetosphere due to a self-polarization electric field. This electric field is set up within the moving plasma clouds by surface charges of opposite signs (electrons and ions) that have been deflected in opposite direction by the Lorentz force and that accumulate on the lateral surfaces of the plasmoids as it proceed across the magnetic field lines with a bulk velocity *u*.

A key parameter for the impulsive penetration mechanism is a kinetic beta:

$$\beta_k = \frac{1}{2}\rho u^2 \bigg/ \frac{B^2}{2\mu_0}, \quad (1)$$

where $\rho$ is the density of plasmoid and *B* is the strength of ambient magnetic field. The excess momentum, i.e. $\beta_k > 1$, results in the drift of plasmoids into the geomagnetic field where they are braked adiabatically. *Ma et al.* [1991] found an additional criterion for the penetration of a plasmoid through the magnetopause: the magnetic fields in a plasmoid and in the magnetosphere should be aligned within ~5°. Otherwise, the kinetic $\beta_k$ should be very large (~50).

*Brenning et al.* [2005] has developed the formalism of impulsive penetration and proposed a scale parameter w':

$$w' = \frac{w}{r_g} k \sqrt{\beta_{th}}, \quad (2)$$

where *w* is the thickness of plasmoid, $r_g$ is the gyro radius of plasmoid ions in the magnetosphere, *k* is an empirical coefficient (*k*~2.3) and $\beta_{th}$ is the ratio of ion thermal pressure in the jet to magnetic pressure in the magnetosphere. Note that the gyro radius $r_g$ is evaluated with the plasma flow velocity. In the two-parametric space, three regimes of plasmoid interaction with the magnetopause were found: rejection ($\beta_{th} < 1$), magnetic expulsion ($\beta_{th} > 1$) and penetration due to self-polarization (*w'* < 1). It was shown that plasmoids with higher $\beta_k$ penetrated more effectively.

*Gunell et al.* [2012] reported a case study of plasma penetration of the high-latitude dayside magnetopause observed by the Cluster mission. They concluded that the conditions for penetration satisfied the mechanism of impulsive penetration. Further from Cluster observations in 2002 - 2006, *Karlsson et al.* [2012] found 16 fast plasmoids with 50% density enhancements in the magnetosheath. Practically all these plasmoids satisfied the conditions for impulsive penetration. From THEMIS observations at low latitudes, *Dmitriev and Suvorova* [2012] found that a high-$\beta$ fast plasmoid (jet) also resulted in magnetosheath plasma transport across the dayside magnetopause.

The identification of magnetosheath plasma penetration inside the magnetosphere meets a problem related to a contribution of the cold plasma of ionospheric and plasmaspheric origin [*Sauvaud et al.*, 2001; *André and Cully*, 2012]. The jet interaction with the magnetopause causes a strong local compression of the geomagnetic field that results in acceleration of the cold plasma with original energy below 5 eV to the energies up to 100 eV. This accelerated cold plasma can mask the plasma of magnetosheath origin. Fortunately, the magnetopause compression/expansion is related to adiabatic acceleration/deceleration of the cold plasma such that a structure of inverted U shape with characteristic duration of few minutes is observed in the ion energy spectrum. On the other hand, the penetrating magnetosheath plasma keeps the energy of ~1 keV for a certain time. This should allow distinguishing between the cold ionospheric and hot magnetosheath plasma.

For the present study, we have collected and analyzed 646 magnetosheath jets observed at low-latitudes by the THEMIS mission in 2007 – 2009. Section 2 describes the technique of jet identification and determination of the magnetosheath plasma penetration of the magnetopause. In Section 2, we also introduce basic macroscopic parameters characterizing the effect of penetration. Conditions for plasma penetration are considered in Section 3. The jet-related magnetosheath plasma flux across the dayside magnetopause is estimated in Section 4. The results are discussed in Section 5. Section 6 is Conclusions.

## 2. Jet identification

Plasma jets were identified with using THEMIS observations of the plasma and magnetic fields in the magnetosheath. The data are freely available at the CDAWeb database (http://cdaweb.gsfc.nasa.gov/). The magnetic field was measured with a time resolution of ~3 sec by the THEMIS/FGM instrument [*Auster et al.*, 2008]. We used high-resolution (~3 sec) plasma data of reduced distribution from the THEMIS/ESA instruments operating in fast survey mode [*McFadden et al.*, 2008]. We also use GOES satellites for analysis of geomagnetic field variations in a geosynchronous orbit.

Experimental data on the upstream solar wind plasma and interplanetary magnetic field were acquired from ACE and Wind monitors rotated around L1 point at geocentric distance of ~230 Earth's radii ($R_E$) upstream of the Earth. Previous studies convincingly showed that the upstream solar wind data acquired near the L1 point are quite reliable for using in the magnetospheric studies [e.g. *Richardson and Paularena*, 2001]. The only crucial problem is accurate determination of the time lag for solar wind propagation from a monitor to the Earth [e.g. *Case and Wild*, 2012].

Figure 1 shows an example of jet identification by using THEMIS data within time interval from 2230 to 2245 UT on 7 August 2007. Table 1 represents the GSM location of the THEMIS probes. They were located in the prenoon sector and slightly southward from the GSM equator. The outermost probe THB was situated most of time in the magnetosheath, a region with a highly variable magnetic field. The innermost probe THA was inside the magnetosphere, which was characterized by a regular geomagnetic field of northward direction. The other probes stay mainly in the magnetosphere.

The upstream solar wind conditions were measured by the ACE monitor. The data on upstream plasma and IMF were provided with 1-min and 15 sec time resolution, respectively. The ACE data were delayed by 43 min for accounting the solar wind propagation time. Note that the solar wind pressure was quasi-stable at that time and, hence, an error in the time delay did not affect the result.

From 2235:45 to 2236:30 UT, the outermost THB observed a fast ($V$ ~ 260 km/s) and dense magnetosheath plasma structure, plasma jet, whose maximum total energy density of ~3 nPa exceeded substantially that one of the incident solar wind (*Ptot* ~ 2.2 nPa). The total energy density, or pressure *Ptot*, is calculated as a sum of the magnetic (*Pm*), the thermal (*Pt*), and the dynamic (*Pd*) pressures. The thermal pressure is the sum of ion and electron thermal pressures. The jet duration of ~45 sec was estimated as the time during which the jet energy density was higher than the solar wind energy density (from 2235:45 to 2236:30 UT). Note that on the time scale of 1-min, the average energy density of the jet is also higher than that of the solar wind. The peak *Ptot* of the jet was contributed by *Pm* ~ 0.4 nPa, *Pt* ~ 1 nPa and *Pd* ~ 1.6 nPa. Hence, the dynamic and thermal energy dominated in the plasma jet.

For a numerical analysis of jet interaction with the magnetopause, we converted the vectors of the THEMIS magnetic field and the plasma velocity into normal coordinates (**l**, **m**, **n**) in the frame of *Lin et al.'s* [2010] reference magnetopause calculated for given upstream solar wind conditions. Here **l** is in the magnetopause plane and points northward; **n** is the magnetopause normal that points outward; and **m** completes the triad by pointing dawnward. As one can see in Figure 1, the components $B_n$ and $B_l$ of the magnetic field measured by THB inside the jet are small and irregular that eliminates reconnection effects such as flux transfer event [*e.g. Elphic*, 1995]. The transversal velocity of the jet corresponds to the plasma flow along the streamlines in the prenoon (dawnward flow) southern (southward flow) sector of the magnetosheath. In addition, the jet has a large normal velocity $V_n$ ~ -70 km/s. It means that the jet is propagating toward the magnetopause and has a chance to interact with it.

The jet interaction with the magnetopause results in local compression of the geomagnetic field [e.g. *Dmitriev and Suvorova*, 2012]. In Figure 1, the compression can be revealed as an increase of the geomagnetic field observed by the THA and THE probes in close vicinity of the magnetopause and by the geosynchronous GOES-11 satellite located at ~13 MLT. Note that the compression was preceded and followed by decreases of geomagnetic field that corresponded to local expansion of the magnetosphere. The expansion can be revealed from the THB encounters with the magnetosphere at 2232 to 2234 UT and at 2240 UT. Hence, the present jet produced a magnetopause distortion of an "expansion – compression – expansion" (ECE) sequence [*Dmitriev and Suvorova*, 2012].

As one can clearly see in Figure 1, the jet resulted in strong magnetopause compression such that the inner probes THC, THD and THE encountered with the magnetosheath from 2236 to 2238 UT. Moreover after returning to the magnetosphere, the probes were observing a portion of the plasma with energies of few keV for ~5 min (from 2237 to 2242 UT). It was unlikely that this plasma belonged to the low-latitude boundary layer (LLBL) because the outer probes THC and THD, located most closely to the magnetopause, did not detect LLBL before the jet as one can see in Figure 1. The plasma, observed by THC, THD and THE, was slightly warmer (few keV) than that observed in the magnetosheath (~1 keV). This might result from conversion of the jet kinetic energy into the thermal one. The energization might also be related to a compression of the magnetic field by the jet. Note that at 2240 UT, the compression was altered by rarefaction, which was accompanied by a decrease of the ion energy as observed by the THE probe.

From the observations presented above, we can qualitatively determine that a large portion of the magnetosheath plasma with energy of a few keV appeared inside the magnetosphere at ~2237 UT and persisted there for ~5 min as one can clearly see in the ion spectra provided by the THC, THD and THE probes. The appearance of the magnetosheath plasma population was accompanied by interaction of the large-scale magnetosheath jet with the magnetopause. Hence, we can suggest that the interaction resulted in penetration of the magnetosheath plasma through the magnetopause into the magnetosphere.

Figure 2 shows another example of plasma penetration through the magnetopause. The penetration resulted from a fast plasma jet observed in the magnetosheath by the THE probe from 1654 to 1656 UT on 6 September 2008. At that time, the probe was continuously located in the prenoon magnetosheath and southward form the GSM equator (see Table 1). In this region, THE observed plasma fluxes along the streamlines directed southward and dawnward. However, the plasma jet had a strong component of velocity toward the magnetopause ($V_n \sim -70$ km/s).

The maximum energy density, measured by the THE probe at 1654:08 UT, was $P_{tot} = 3.7$ nPa and it was contributed by $P_t \sim 2.1$ nPa, $P_d \sim 1$ nPa and $P_m \sim 0.5$ nPa. The energy density of the high-beta fast plasma jet was 1.44 times higher than the energy density of incident solar wind ($P_{tot} = 2.6$ nPa) measured 54 minutes earlier by the ACE upstream monitor. The time delay for the solar wind propagation was obtained from cross-correlation between the magnetic field components measured by ACE in the upstream solar wind and by THE in the magnetosheath. Note that THEMIS probes THB, THC and THD, located upstream of the bow shock, were also observing the IMF components very similar to those observed by ACE.

As one can see in Figure 2, the jet was preceded by 1-hour interval of multiple magnetopause crossings by the THE probe. The interval was also characterized by multiple intensifications of the low-energy plasma in the noon magnetosphere as observed by the THA probe. Those intensifications could be produced by the accelerated cold plasma and/or by the plasma penetrated from the magnetosheath. As a result, the THA probe observed a mixture of hot magnetospheric ions with energies above several keV and low-energy ions with energies from several hundreds keV to few keV originated from the cold magnetospheric plasma population and from the magnetosheath.

At the time of jet observation by the THE probe from ~1653 to 1656 UT, the THA probe detected a prominent enhancement of the low-energy ions in the magnetosphere. The enhancement was accompanied by a compression of the geomagnetic field as observed by THA from 1652 to 1657 UT and by GOES-12 at geosynchronous orbit from 1653 to 1659 UT. Note that the magnetic variations indicate the ECE sequence in the magnetopause distortion. Apparently, the compression resulted from the jet interaction with the magnetopause.

This case event is different from the previous one because the magnetosheath plasma was persisting in the magnetosphere after the previous disturbances. In the present case, we can qualitatively identify the plasma penetration as a prominent enhancement of ~1 keV ions observed by the THA probe inside the magnetosphere during interaction of the magnetopause with the plasma jet observed by the THE probe in the magnetosheath. Hence, we can suggest that the jet resulted in penetration of the magnetosheath plasma through the magnetopause that provided a fresh portion of ~1 keV ions inside the magnetosphere.

Figure 3 shows an example of jet, which does not result in plasma penetration. The plasma jet was observed in the magnetosheath by the THA probe from 0422:55 to 0423:25 UT on 5 September 2007. At that time, THEMIS probes were located in the prenoon sector (~11 MLT) and slightly southward from the GSM equator (see Table 1). The peak energy density of the jet of 3.1 nPa was 1.7 times higher than that of the solar wind observed by the ACE upstream monitor. The time delay of 49 min for the solar wind propagation was obtained from cross-correlation between the magnetic components measured by ACE and THA.

From 0447 to 0432 UT, the THA probe was continuously located in the magnetosheath, where it was detecting plasma streams directed dawnward ($V$m > 0) and southward ($V$l < 0) in according to the streamlines. The jet was characterized by a relatively high speed of ~285 km/s (versus ~100 km/s for the magnetosheath plasma) with a strong component $V$n = -240 km/s directed toward the magnetopause. The jet kinetic energy density $P$d = 2.2 nPa was much higher than the thermal $P$t = 0.74 nPa and magnetic $P$m = 0.18 nPa energy density. Hence from the THA data, we can identify a fast high-$\beta$ plasma jet. Note that while $B$l was negative, the jet could not be related to magnetic reconnection because the normal component of magnetic field $B$n was tenuously small.

The fast plasma jet interacted with the magnetopause that resulted in a magnetopause compression observed by THD and THC from 0422:10 to 0425:25 UT as magnetosheath encounters due to inward motion of the magnetopause. The compression can also be revealed from an increase of the geomagnetic field detected by the innermost THA and THC probes from 0421:30 to ~0426 UT. Note that geosynchronous satellite GOES-11, located in the dusk sector at ~1850 MLT, detected a ECE sequence from 0418:30 to 0427:30 UT. The observed dynamics of the magnetopause and geomagnetic field indicate a substantial distortion of the magnetopause by the jet.

However in contrast to the cases presented above, the innermost probes THA and THC did not detect any remnants of the magnetosheath plasma inside the magnetosphere. Instead, they observed a plasma structure of inverted U shape that corresponded to acceleration of the cold plasma from original energy below 5 eV to several hundreds of eV and following plasma deceleration to the original energy.

Figure 4 demonstrates a very strong jet interacting with the magnetopause and not resulting in plasma penetration. The jet was detected by all THEMIS probes within time interval from 1048 to 1053 UT on 21 July 2007. At that time, THEMIS probes were located in the postnoon sector (~13 MLT) at low latitudes in the Southern GSM Hemisphere. The outermost probe THA was located in the magnetosheath and observed the jet from 1048:36 to 1052:40 UT. The peak energy density of the jet was $P\text{tot} = 6.1$ nPa while the energy density of the upstream solar wind, observed 50 min earlier by ACE, was 3.4 times smaller. The jet was very fast ($V = 360$ km/s) such that its kinetic energy density of $P\text{d} = 4.3$ was much higher that the thermal $P\text{t} = 1.2$ nPa and magnetic $P\text{m} = 0.6$ nPa energy densities. Transversal plasma stream in the jet was oriented along the magnetosheath streamlines toward south ($V\text{l} < 0$) and dusk ($V\text{m} < 0$). In addition, the jet had very high normal component of the velocity $V\text{n} = -270$ km/s toward the magnetopause. Significant earthward inclination of the jet from the magnetosheath streamlines produced a large magnetic component $B\text{n}$ (normal to the magnetopause) observed by THA inside the jet.

Interaction of the fast high-$\beta$ plasma jet with the magnetopause resulted in very strong distortion and compression of the geomagnetic field. The magnetopause moved inward on at least 0.7 $R_E$ from THE to THB. The innermost probe THB observed a strong increase of geomagnetic field from 36 to 82 nT (2.28 times) that was interrupted by a brief encounter with the magnetosheath at 1049:55 - 1051:30 UT. Note that the number of 2.28 is quite close to the theoretical prediction of 2.44 for the magnification of the dipole magnetic field by shielding currents at the magnetopause [*e.g. Shield*, 1969; *Shabansky*, 1972]. At geosynchronous orbit, the GOES-12 satellite detected a 3 nT increase of the geomagnetic field in the dawn sector at 0510MLT. Despite of the very strong distortions, magnetosheath plasma did not penetrate across the magnetopause and was not observed inside the magnetosphere by the THB, THC and THD probes.

It is interesting to note a fast sunward plasma flow with $V$n > 200 km/s observed by the THA probe from 1048:10 to 1049:05 UT. The sunward flow was also observed by other probes (not shown). The innermost THB detected the plasma flow with $V$n ~ 150 km/s from 1049:50 to 1050:45 UT. It is reasonable to suggest that this outflow is due to deflection of the plasma jet by the magnetopause. Note that similar phenomenon was reported by *Shue et al.* [2009]. Hence in the present case, the plasma of jet was deflected from the magnetopause rather than penetrated across it.

## 3. Conditions for penetration

Using the method described above, we have collected 646 large-scale plasma jets observed in the magnetosheath by THEMIS from 2007 to 2009. For the jets, the ratio $R$ of total energy density of a jet to that of the upstream solar wind was required to be larger than 1 for $\Delta T$ = 30 seconds and longer. The duration of jet $\Delta T$ was determined when both $R > 1$ and $V$n < 0. Another important criterion of the selection was a compression of the geomagnetic field that indicated an interaction of a jet with the magnetopause. The spatial range of jet location was restricted by 80° in longitude in order to avoid an effect of plasma penetration due to K-H waves at the magnetopause flank and tail. Figure 5 shows the GSM location of the jets. Because the THEMIS orbit specifics, the statistics of jets in the Southern Hemisphere prevails over the Northern one and the vast majority of the jets are found at low latitudes. Apparently, not every jet, observed by a THEMIS probe in the magnetosheath, was accompanied by simultaneous measurements by another probe in the magnetosphere. Hence, an additional selection has been conducted in order to analyze the conditions for the magnetosheath plasma penetration of the magnetopause. The selection was based on a criterion that the magnetosheath plasma penetration or not penetration into the magnetosphere could be determined unambiguously. Namely, one THEMIS probe should provide high-resolution data on plasma and magnetic field in the magnetosheath and another – in the magnetosphere in a close vicinity (~1 $R_E$) of the magnetopause. While a jet is observed in the magnetosheath, a probe located in the magnetosphere should either observe or not observe the population of magnetosheath plasma.

As a result, we have selected 76 jets. Table 2 lists the peak time of the jets and the probes used for the jet identification. The duration of jets varies from 0.5 to 3 min with the most probable mean of 1 min. We have found that 44 out of 76 jets (almost 60%) were accompanied by the penetration (hereafter penetrating jets). The other 32 jets were not accompanied by the penetration (hereafter nonpenetrating jets). A decrease of statistics from 2007 to 2009 is explained by an increase of the THEMIS apogee on the dayside such that the probes spend less time in the magnetosheath. In Figure 5, one can see that the location of both kinds of the jets in GSM coordinates is scattered quite randomly. Hence, the penetration does not depend on the jet location.

Figure 6 shows a scatter plot of the normal component of the peak jet velocity $V_n$ versus the ratio $R$ of the total energy densities. One can see that jets with large $R$ and high $V_n$ are mainly the penetrating jets. However, there are many penetrating jets with small $R$ and low $V_n$. The nonpenetrating jets are scattered quite randomly. Hence, the penetration conditions are controlled neither by the ratio $R$ of total energy densities nor by the plasma velocity $V_n$ normal to the magnetopause.

Figure 7 demonstrates a scatter plot of the geomagnetic field magnitude $B_{mp}$ at the magnetopause versus the north-south component $B_l$ of the jet magnetic field. The geomagnetic field at the magnetopause was calculated from the following formula:

$$B_{mp} = 2.44 \cdot B_d. \qquad (3)$$

Here $B_d$ is the dipole magnetic field calculated from IGRF model of epoch 2005 for the geocentric distance of the reference magnetopause, which was calculated from the *Lin et al.'s* [2010] model for the angular coordinates of a jet. The coefficient 2.44 is acquired from a self-consistent solution of the Chapman-Ferraro problem at the subsolar magnetopause [*Mead*, 1964].

In Figure 7, one can clearly see that the magnetosheath plasma might penetrate through the magnetopause and, hence, might appear in the outer magnetosphere under strong magnetic field at the magnetopause (up to 90 nT). And vice versa, the magnetosheath plasma might not penetrate through the magnetopause characterized by relatively weak geomagnetic field of ~40 nT. Both kinds of the jets are also distributed randomly in the $Bl$ component of the magnetosheath magnetic field. It is important to note that the $Bl$ component controls magnetic reconnection at the magnetopause: magnetosheath magnetic field with negative $Bl$ can reconnect with the strong positive $Bl$ dominating in the dayside low-latitude magnetosphere. Both penetrating and nonpenetrating jets were observed under negative and positive values of $Bl$. Hence, the magnetic reconnection plays a minor role (if any) in the plasma penetration related to the large-scale plasma jets.

Figure 8 shows a distribution of the jets in a space of parameters $\beta_k$ (see Eq. 1) and $w'$ (see Eq. 2). In calculation of the scale parameter $w'$, we used the following simplifications. The thickness of jet $w$ is calculated as:

$$w = V_t \cdot \Delta T, \quad (4)$$

where $V_t$ is the transversal velocity along the magnetopause and $\Delta T$ is the duration of a jet. The average thickness of jets was found to be ~2 Re. Note that this method does not take into account the orientation of a jet relative to its propagation along the magnetopause that leads to overestimation of the thickness. The gyro radius $r_g$ of the jet ions in the magnetosphere was calculated using the following expression:

$$r_g = m_p V / eB_{mp}. \quad (5)$$

Here $m_p$ and $e$ are the mass and electric charge of proton, $V$ is the bulk velocity of a jet and $B_{mp}$ is the geomagnetic field at the magnetopause (see Eq. 3). Here we assume that the jet characteristics do not change much during its propagation across the magnetosheath to the magnetopause.

As one can see in Figure 8, the scale parameter $w'$ varies from several tens to a few thousands. That is in good agreement with the results reported by *Karlsson et al.* [2012]. This range of *w'* corresponds to the regimes of rejection ($\beta_k < 1$) and magnetic expulsion ($\beta_k > 1$) separated by a boundary of $\beta_k = 1$. In Figure 8, we find that indeed all the penetrating jets are characterized by the kinetic $\beta_k$ varying from 1 to ~20. The nonpenetrating jets are characterized by the kinetic $\beta_k$ varying from 0.2 to ~5 such that $\beta_k > 1$ for many of them. Hence, the boundary of $\beta_k = 1$ does not firmly separate the penetrating and nonpenetrating jets.

**4. Plasma flux across the magnetopause**

In order to estimate the jet-related plasma flux across the magnetopause, we have to find a parametric space, in which the penetrating and nonpenetrating jets can be separated more or less reliably. We analyzed various couples of parameters and found that the best separation can be obtained in the parametric space of jet velocity *V* and kinetic $\beta_k$ as shown in Figure 9. Note that the plasma velocity *V* was available only for 472 out of 646 large-scale plasma jets.

As one can see in Figure 9, the vast majority of penetrating jets (41 out of 44, i.e. 93%) are characterized by $\beta_k > 1$ and $V > 220$ km/s. The nonpenetrating jets are mainly characterized either by $\beta_k < 1$ or by low velocities ($V < 220$ km/s). Such a separation is reasonable because the jets with lower velocities bring ions with smaller gyro radii, which have less capability for penetration through the magnetic barrier of the magnetopause. However, we find 7 out of 32 (i.e. ~20%) nonpenetrating jets, which are characterized by both kinetic $\beta_k > 1$ and high velocity ($V > 250$ km/s). One of such jets was presented in Figure 4 (see Section 2).

The parametric space of *V* and $\beta_k$ allows separation of 93% of the penetrating jets from 80% of the nonpenetrating ones. We can apply these criteria to the statistics of 472 jets in order to determine the plasma fluxes of penetrating jets. We found that 273 out of 472 jets (i.e. ~60%) satisfy the conditions of $\beta_k > 1$ and $V > 220$ km/s. The plasma flux *F* across the magnetopause can be defined as:

$$F = nV_n, \quad (6)$$

where $n$ and $V_n$ are, respectively, plasma density and velocity component normal to the nominal magnetopause in the maximum of a jet. It is reasonable to assume that the flux of penetrating plasma should be transversal to the magnetosheath streams and normal to the magnetopause. Figure 10 shows statistical distribution of the peak plasma fluxes $F$ in 273 large-scale jets, which satisfy the penetration conditions. The fluxes vary from $10^6$ to $10^9$ (cm$^2$ s)$^{-1}$ with the most probable mean value of $F = 3*10^8$ (cm$^2$ s)$^{-1}$. Note that the average flux of the solar wind plasma with typical density of 5 cm$^{-3}$ and velocity of 400 km/s is equal to 2 $10^8$ (cm$^2$ s)$^{-1}$, i.e. 1.5 times less.

The total amount of magnetosheath plasma $I$ penetrating with a jet through the magnetopause can be estimated from the following expression:

$$I = F \cdot \Delta T \cdot l^2, \qquad (7)$$

where $\Delta T$ and $l$ are, respectively, the duration and size of a jet. The duration $\Delta T$ was defined in the beginning of Section 3 and varied from jet to jet. In a first approach, we can replace the size of a jet by its thickness, i.e. $l = w$. Figure 11 shows the statistical distribution of the total amount of plasma penetrating to the magnetosphere with jets. The amount varies from $5*10^{26}$ to $2*10^{29}$ particles with the most probable mean of $5*10^{28}$ particles. Vast majority of the jets brings more than $10^{28}$ particles across the magnetopause.

Further, we can estimate an average rate of jet-related magnetosheath plasma penetration through the dayside magnetopause. During summer months of the year 2007, THEMIS was located outside the magnetosphere, predominantly in the magnetosheath, for about 18 hours per day. During this time, we have identified 106 penetrating jets. This statistics corresponds to the occurrence frequency of almost 2 jets per day. Hence, the total amount of plasma carrying by the large-scale jets into the low-latitude dayside magnetosphere can be estimated to be ~$10^{29}$ particles per day.

## 5. Discussion

Apparently, the plasma penetration rate of ~$10^{29}$ particles per day is only a rough estimation. Usually,

the jets do not occur every day. Sometimes, several jets can be found within a few hours. For example on 7 August 2007, five jets occurred within 2 hours from 9 to 11 UT and the total amount of penetrated plasma constituted of ~$3*10^{29}$ particles that corresponded to the rate of $1.5*10^{29}$ particles per hour (i.e. $4*10^{25}$ ions/s). This rate is close to estimates of the total amount of plasma entering the dayside magnetosphere that is on the order of $10^{26}$ ions/s [*Sibeck*, 1999]. It is also comparable with the dayside outflow of low-energy ions with energies below tens of eV, which vary from a few times $10^{25}$ ions/s to $10^{27}$ ions/s in plasmaspheric plumes [*André and Cully*, 2012]. However, the plasma of magnetosheath jets is much hotter (~1 keV) and can contribute effectively to the formation of LLBL.

Our estimation of the jet-related plasma flux across the magnetopause is based on a number of assumptions and simplifications. First of all, we assumed that a jet is not changed much during propagation across the magnetosheath toward the magnetopause. However, that is not completely correct. Figure 12 and Table 3 demonstrate a dynamics of the key parameters obtained at different geocentric distances by the THEMIS probes for a large-scale magnetosheath plasma jet observed around 0814 UT on 23 June 2007. One can see that in the outer region of the subsolar magnetosheath at distance of 12.3 $R_E$, the outermost probe THB observed a long jet ($\Delta T = 100$ s) with moderate values of parameters: the bulk and normal velocity, respectively, $V = 270$ km/s and $V_n = -160$ km/s, the energy density ratio $R = 1.58$ and the kinetic $\beta_k = 1.33$. At the same time, the THC probe at distance of 11.8 $R_E$ observed a well-developed jet of 70 sec duration. With approaching to the magnetopause, the jet was slowing down from 380 km/s to 320 km/s and its duration was increasing from 70 to 80 sec. The strength of jet expressed in $R$ and $\beta_k$ was decreasing with the decreasing distance.

It is interesting to note that the magnitude of normal velocity $V_n$ of the jet was increasing from –280 km/s to –320 km/s as observed by the probes THC, THD, THE and the innermost probe THA located in close vicinity of the magnetopause. It looked like the jet was focused during its propagation across the magnetosheath toward the magnetopause. Hence, the characteristics of jet can vary significantly with the jet location in the magnetosheath. This effect resulted in a decrease of the number of jets identified because sometimes the information about plasma velocity was not available from some of THEMIS probes. Therefore, the rate of jet-related magnetosheath plasma penetration across the dayside magnetopause might be higher.

Considering the conditions for jet penetration, we simplified the calculation of jet thickness with using Equation 4. Unambiguous determination of the thickness is difficult because the THEMIS probes are mainly stretched radially that makes the triangulation unreliable. Moreover, accurate determination of the thickness affects only the scale parameter *w'*, which does not control the penetration conditions in the range of *w'* > 1 [*Brenning et al.*, 2005]. More serious assumption concerns to estimation of the jet scale. Using Equation 4, we have obtained that the average scale of jets is ~2 $R_E$. On the other hand, *Dmitriev and Suvorova* [2012] reported a plasma jet whose scale was estimated to be ~8 $R_E$. Hence, the actual size of large-scale plasma jets can be larger than 2 $R_E$ that might result in a few time increase of the penetration rate.

The flux of magnetosheath plasma across the magnetopause was determined from Equation 6. As we showed above, the normal component of jet velocity $V_n$ can increase with approaching to the magnetopause that, in turn, increases the penetrating flux *F*. On the other hand, the plasma flux might be deflected in interaction with the magnetopause [e.g. *Shue et al.*, 2009; *Dmitriev and Suvorova*, 2012] that results in a decrease of the penetrating plasma flux. This effect leads to a decrease of the penetration rate. Estimation of the penetration efficiency is a subject of further studies.

Finally, an additional decrease of the penetration rate is originated from imperfect criteria developed for separation between the penetrating and nonpenetrating jets. The threshold of *V* = 220 km/s was obtained empirically on the base of limited statistics. The existence of such threshold can be explained from the finite Larmor radius effect. Namely, lower velocity of a jet leads to smaller gyro radius of the ions as follows from Equation 5. The low-latitude dayside magnetopause has a thickness of ~500 km. In order to penetrate through this barrier, the ions should have sufficiently large gyro radius and, thus, a jet should be fast enough to carry such ions. Hence, the threshold velocity of ~220 km/s corresponds to the average thickness of the magnetopause.

Another shortcoming of the separation criteria is a presence of fast and high $\beta_k$ nonpenetrating jets, which constitute ~20% of the statistics of nonpenetrating jets. The origin of such kind of jets is unknown and should become a subject of further studies. The nonpenetrating jets contribute 40% of the whole statistics, while the penetrating jets contribute the other 60%. Hence, the false plasma rate across the magnetopause, related to the fast and high $\beta_k$ nonpenetrating jets, can be estimated to be 0.4 * 0.2 = 8% that leads to ~10% uncertainty in determination of the plasma rate.

**6. Conclusions** From THEMIS observations of 642 large-scale plasma jets in the low-latitude magnetosheath and their interaction with the magnetopause, we have found the following:

1. About 60% of the jets are accompanied by magnetosheath plasma penetration through the magnetopause into the magnetosphere.

2. Vast majority of the penetrating jets is characterized by high velocities $V > 220$ km/s and kinetic $\beta_k > 1$ that corresponds to a combination of the finite Larmor radius effect with the mechanism of impulsive penetration.

3. The average plasma flux in the penetrating jets is estimated to be $\sim 3*10^8$ $(cm^2\ s)^{-1}$ that is 1.5 times larger than the average plasma flux of the solar wind.

4. The average rate of jet-related penetration of the magnetosheath plasma into the magnetosphere was obtained to be $\sim 10^{29}$ particles per day. The rate varied highly from day to day and sometime achieved values of $1.5*10^{29}$ particles per hour that is comparable with estimates of the total amount of plasma entering the dayside magnetosphere.

**Acknowledgements** The authors are grateful CDAWeb (http://cdaweb.gsfc.nasa.gov/) for providing free public data on magnetic field and plasma measured by the THEMIS mission. We acknowledge NASA contract NAS5-02099 and V. Angelopoulos for use of plasma data from the THEMIS mission. We thank K.H. Glassmeier and U. Auster for the use of magnetic FGM data provided under contract 50 OC 0302 and C. W. Carlson and J. P. McFadden for use of ESA data. We thank N. Ness and D.J. McComas for the use of ACE solar wind data made available via the CDAWeb. The authors are grateful Howard J. Singer for the opportunity to use magnetic data from the GOES geosynchronous satellites presented at CDAWeb. We appreciate J.-K. Chao and S. Savin for the useful discussion and comments. This work was supported by grant NSC-102-2111-M-008-023- from the National Science Council of Taiwan and by Ministry of Education under the Aim for Top University program at National Central University of Taiwan.

**Table 1.** GSM location (geocentric distance, local time and latitude) of THEMIS probes at time of jet observations

| Date and time | THA | THB | THC | THD | THE | PI* |
|---|---|---|---|---|---|---|
| 21 Jul 2007 1051UT | 12.7 $R_E$ 1250MLT -21.7° | 11.0 $R_E$ 1320MLT -23.7° | 11.5 $R_E$ 1305MLT -23.2° | 11.5 $R_E$ 1305MLT -23.1° | 11.7 $R_E$ 1305MLT -23.2° | THA |
| 7 Aug 2007 2236UT | 8.6 $R_E$ 0920MLT -10.3° | 11.0 $R_E$ 1000MLT -12.5° | 10.5 $R_E$ 0950MLT -12.1° | 10.4 $R_E$ 0950MLT -12.0° | 10.2 $R_E$ 0950MLT -12.2° | THB |
| 5 Sep 2007 0423UT | 11.9 $R_E$ 1035MLT -8.5° | 9.9 $R_E$ 1105MLT -10.5° | 10.0 $R_E$ 1105MLT -10.2° | 10.4 $R_E$ 1100MLT -9.7° | 10.4 $R_E$ 1100MLT -9.9° | THA |
| 6 Sep 2008 1654UT | 10.2 $R_E$ 1220MLT -15.4° | 29.9 $R_E$ 0940MLT -4.1° | 16.8 $R_E$ 1130MLT -6.7° | 11.5 $R_E$ 1120MLT -10.1° | 11.1 $R_E$ 1055MLT -10.2° | THE |

*PI - probe observing the jet

**Table 2.** List of large-scale plasma jets identified by THEMIS in the magnetosheath

| year | mon | day | UT | R | PI | year | mon | day | UT | R | PI |
|------|-----|-----|------|------|------|------|-----|-----|------|------|------|
| 2007 | 5 | 8 | 1522 | 1.98 | *THC**  | 2007 | 8 | 8 | 1642 | 2.33 | *THA* |
| 2007 | 5 | 8 | 1550 | 1.10 | **THC#** | 2007 | 8 | 11 | 817 | 2.16 | *THC* |
| 2007 | 5 | 24 | 647 | 1.88 | **THB** | 2007 | 8 | 11 | 2235 | 2.25 | **THB** |
| 2007 | 5 | 26 | 2243 | 2.68 | **THB** | 2007 | 8 | 20 | 1035 | 2.04 | *THA* |
| 2007 | 6 | 3 | 1713 | 1.90 | **THC** | 2007 | 8 | 29 | 1531 | 1.99 | *THA* |
| 2007 | 6 | 9 | 1408 | 1.29 | **THA** | 2007 | 8 | 29 | 1542 | 1.37 | *THA* |
| 2007 | 6 | 9 | 1419 | 1.36 | **THA** | 2007 | 9 | 3 | 405 | 2.66 | **THD** |
| 2007 | 6 | 15 | 1032 | 2.64 | **THB** | 2007 | 9 | 3 | 412 | 1.54 | **THD** |
| 2007 | 6 | 15 | 1039 | 2.99 | **THB** | 2007 | 9 | 4 | 1237 | 2.29 | **THD** |
| 2007 | 6 | 15 | 1055 | 1.39 | *THD* | 2007 | 9 | 5 | 413 | 2.49 | *THA* |
| 2007 | 6 | 15 | 1101 | 1.76 | *THB* | 2007 | 9 | 5 | 423 | 1.77 | *THA* |
| 2007 | 6 | 16 | 454 | 2.49 | **THE** | 2008 | 6 | 25 | 2026 | 2.12 | **THD** |
| 2007 | 6 | 18 | 312 | 1.82 | *THB* | 2008 | 7 | 11 | 2047 | 2.98 | **THD** |
| 2007 | 6 | 18 | 1835 | 1.62 | **THA** | 2008 | 7 | 23 | 1633 | 2.56 | *THD* |
| 2007 | 7 | 1 | 1906 | 1.76 | **THA** | 2008 | 8 | 9 | 1809 | 3.03 | **THE** |
| 2007 | 7 | 4 | 1208 | 1.75 | **THA** | 2008 | 8 | 9 | 1834 | 1.91 | **THE** |
| 2007 | 7 | 4 | 1214 | 1.77 | **THE** | 2008 | 8 | 18 | 1644 | 2.86 | *THE* |
| 2007 | 7 | 5 | 059 | 1.15 | *THB* | 2008 | 8 | 29 | 2311 | 3.72 | *THC* |
| 2007 | 7 | 5 | 111 | 1.57 | *THB* | 2008 | 9 | 4 | 1634 | 1.46 | *THD* |
| 2007 | 7 | 10 | 601 | 2.72 | *THB* | 2008 | 9 | 4 | 1843 | 2 | **THE** |
| 2007 | 7 | 12 | 2034 | 2.72 | **THB** | 2008 | 9 | 5 | 1858 | 1.59 | *THD* |
| 2007 | 7 | 18 | 247 | 2.12 | *THB* | 2008 | 9 | 6 | 1654 | 1.44 | **THE** |
| 2007 | 7 | 18 | 311 | 2.92 | *THB* | 2008 | 9 | 8 | 1614 | 1.41 | *THD* |
| 2007 | 7 | 18 | 321 | 2.24 | *THB* | 2008 | 9 | 15 | 1623 | 2.87 | **THE** |
| 2007 | 7 | 21 | 1051 | 3.44 | *THA* | 2008 | 9 | 30 | 2243 | 2.56 | **THE** |
| 2007 | 7 | 29 | 2007 | 1.16 | *THE* | 2008 | 10 | 4 | 1801 | 4.49 | **THE** |
| 2007 | 8 | 2 | 428 | 3.76 | **THA** | 2008 | 10 | 11 | 1521 | 2.36 | **THE** |
| 2007 | 8 | 7 | 945 | 1.79 | **THA** | 2008 | 10 | 11 | 1528 | 1.6 | **THE** |
| 2007 | 8 | 7 | 1035 | 1.37 | *THC* | 2008 | 10 | 11 | 1548 | 1.56 | **THD** |
| 2007 | 8 | 7 | 1048 | 2.14 | *THA* | 2008 | 10 | 22 | 1610 | 2.87 | **THE** |
| 2007 | 8 | 7 | 1056 | 4.89 | **THA** | 2008 | 10 | 22 | 2128 | 1.08 | *THA* |
| 2007 | 8 | 7 | 2225 | 4.24 | **THB** | 2008 | 10 | 30 | 1432 | 1.24 | **THD** |
| 2007 | 8 | 7 | 2236 | 1.45 | **THB** | 2008 | 6 | 25 | 1544 | 2.32 | **THA** |
| 2007 | 8 | 7 | 2258 | 2.98 | **THB** | 2008 | 8 | 12 | 010 | 3.77 | **THE** |
| 2007 | 8 | 7 | 2303 | 2.12 | **THC** | 2008 | 9 | 15 | 1605 | 1.5 | *THD* |
| 2007 | 8 | 7 | 2358 | 2.08 | **THB** | 2008 | 9 | 15 | 1621 | 2.78 | **THE** |
| 2007 | 8 | 8 | 011 | 1.64 | *THD* | 2008 | 9 | 24 | 1354 | 1.43 | *THD* |
| 2007 | 8 | 8 | 1637 | 2.23 | *THE* | 2009 | 7 | 12 | 2246 | 4.26 | **THC** |

*Italic font for penetrating jets
#Bold font for nonpenetratig jets

**Table 3.** Radial dynamics of a large-scale magnetosheath plasma jet observed by THEMIS around 0814 UT on 23 June 2007.

| Probe | Rgc*, $R_E$ | MLT | GSM Lat | $R$ | $\beta_k$ | $V$, km/s | $V$n, km/s | $\Delta T$, s |
|---|---|---|---|---|---|---|---|---|
| THB | 12.3 | 1250 | -16.6 | 1.58 | 1.33 | 270 | -160 | 100 |
| THC | 11.8 | 1242 | -16.2 | 2.88 | 3.11 | 380 | -280 | 70 |
| THD | 11.8 | 1240 | -16.2 | 2.50 | 2.61 | 360 | -290 | 70 |
| THE | 11.6 | 1239 | -16.2 | 2.53 | 2.74 | 370 | -320 | 75 |
| THA | 10.4 | 1225 | -14.8 | 2.23 | 2.56 | 320 | -320 | 80 |

*Geocentric distance

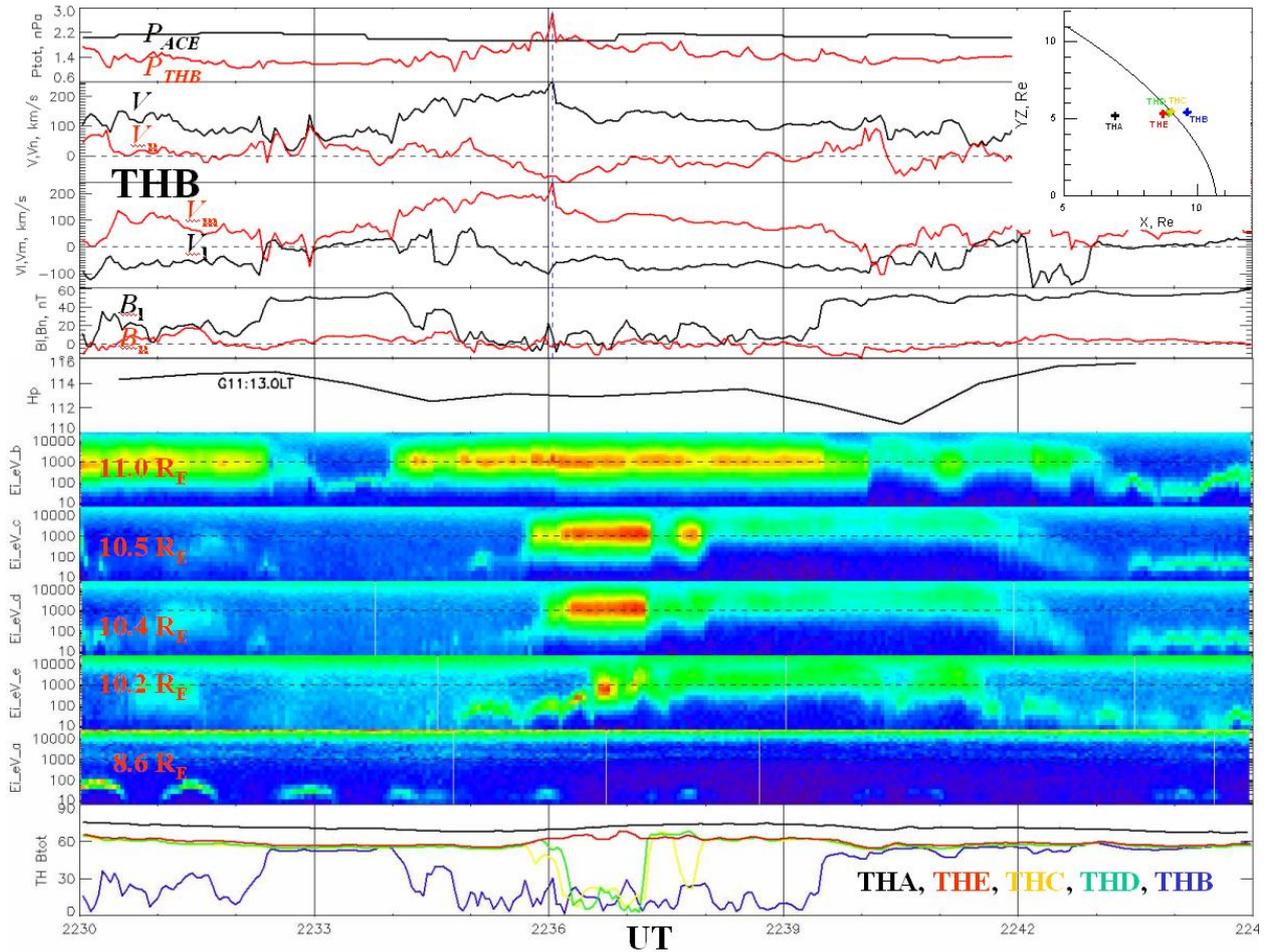

Figure 1. Observations of plasma and magnetic field on 7 August 2007 (from top to bottom): total energy density measured by the ACE upstream monitor (black curve) and THB probe (red curve); bulk velocity *V* (black curve) and its component *V*n (red curve) normal to *Lin et al.'s* [2010] nominal magnetopause calculated from THB plasma measurement and ACE upstream conditions; transversal components *V*l (black curve) and *V*m (red curve) of plasma velocity measured by the THB probe (see details in the text); component *B*l (black curve) and *B*n (red curve) of magnetic field measured by THB; horizontal magnetic field detected by the GOES-11 geosynchronous satellite at 1300MLT; ion spectrograms measured by THB, THC, THD, THE and THA; magnitude of magnetic field measured by THA, THE, THC, THD and THB shown, respectively, by black, red, yellow, green and blue curves. The ACE measurements are delayed by 43 min. Red numbers indicate the geocentric distance of THEMIS probes. From 2235:45 to 2236:30 UT the THB observed a fast magnetosheath plasma jet. From 2227 to 2242, the probes THC, THD and THE observed the magnetosheath plasma inside the magnetosphere. Panel in the upper right corner shows the location of THEMIS probes and *Lin et al.'s* [2010] magnetopause in GSM coordinates.

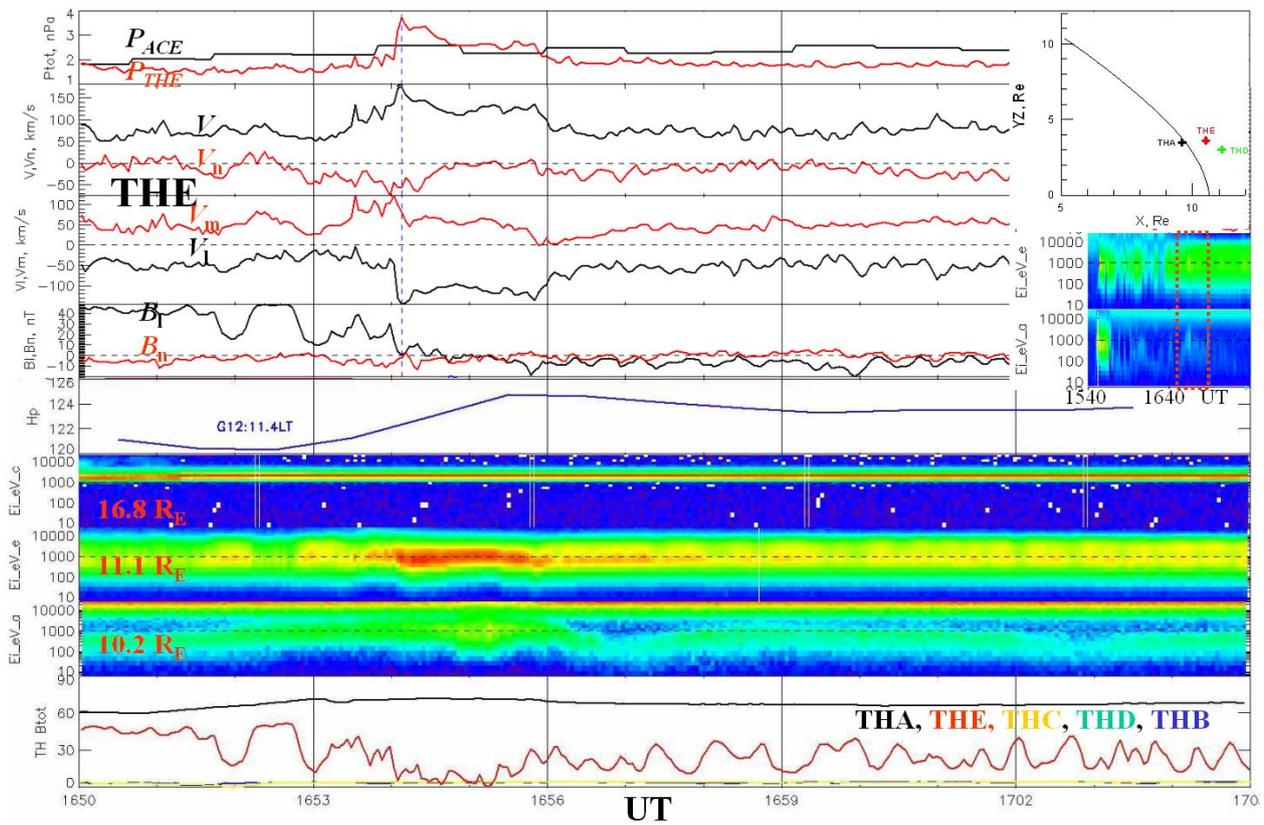

Figure 2. The same as in Figure 1 but for the THE probe on 6 September 2008. The ACE upstream data were delayed by 54 minutes. The horizontal magnetic field at geosynchronous orbit was observed by GOES-12 at 1125 MLT. Ion spectra were not available from the probes THB and THD. A fast magnetosheath plasma jet was observed from 1654 to 1656 UT. The penetration of magnetosheath plasma was observed inside the magnetosphere by the THA probe from 1653 to 1656 UT. For reference, the ion spectra measured from 1540 to 1740 UT by the THE and THA probes are shown on the right side.

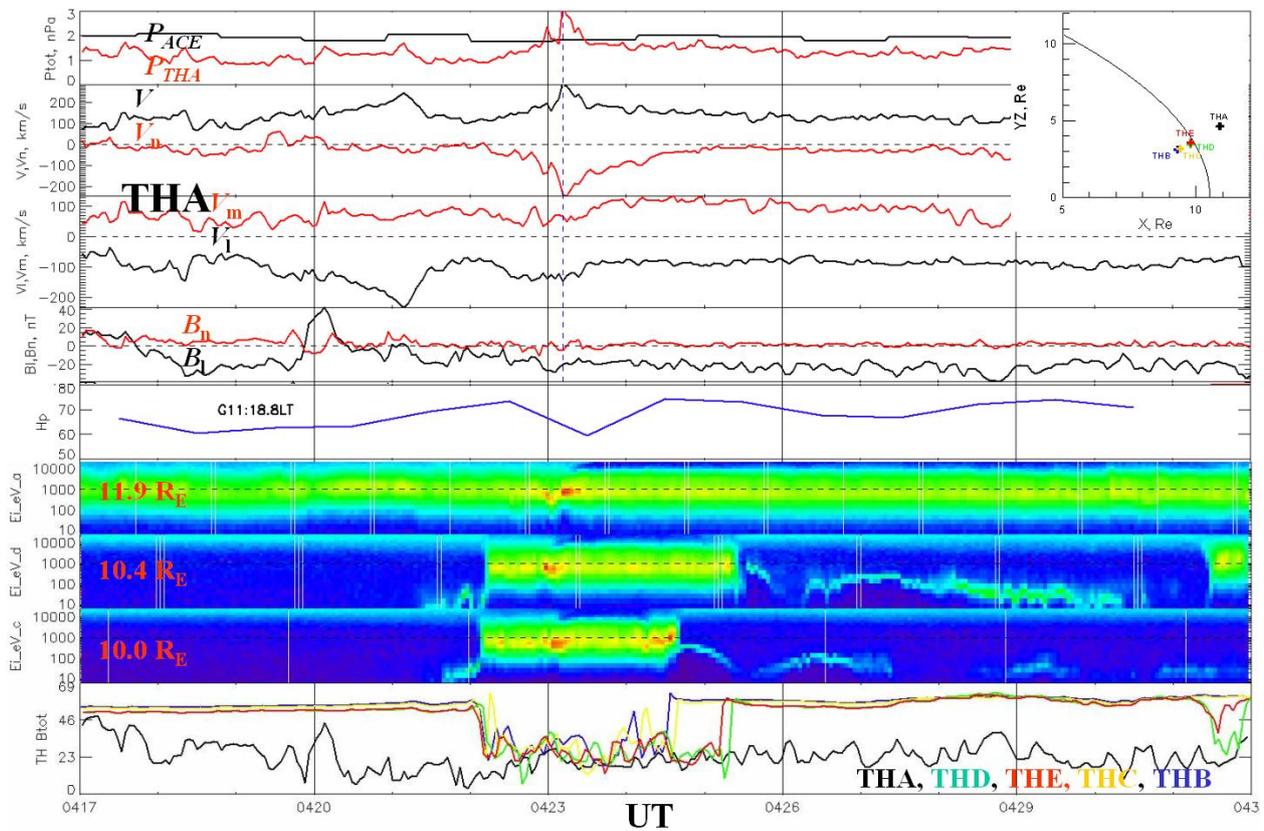

Figure 3. The same as in Figure 1 but for the THA probe on 5 September 2007. The ACE upstream data were delayed by 49 minutes. The magnetic field at geosynchronous orbit was observed by the GOES-11 satellite at 1850MLT. A fast magnetosheath plasma jet was observed by the THA probe from 0422:55 to 0423:25 UT. Both THD and THC did not observe any portion of the magnetosheath plasma inside the magnetosphere.

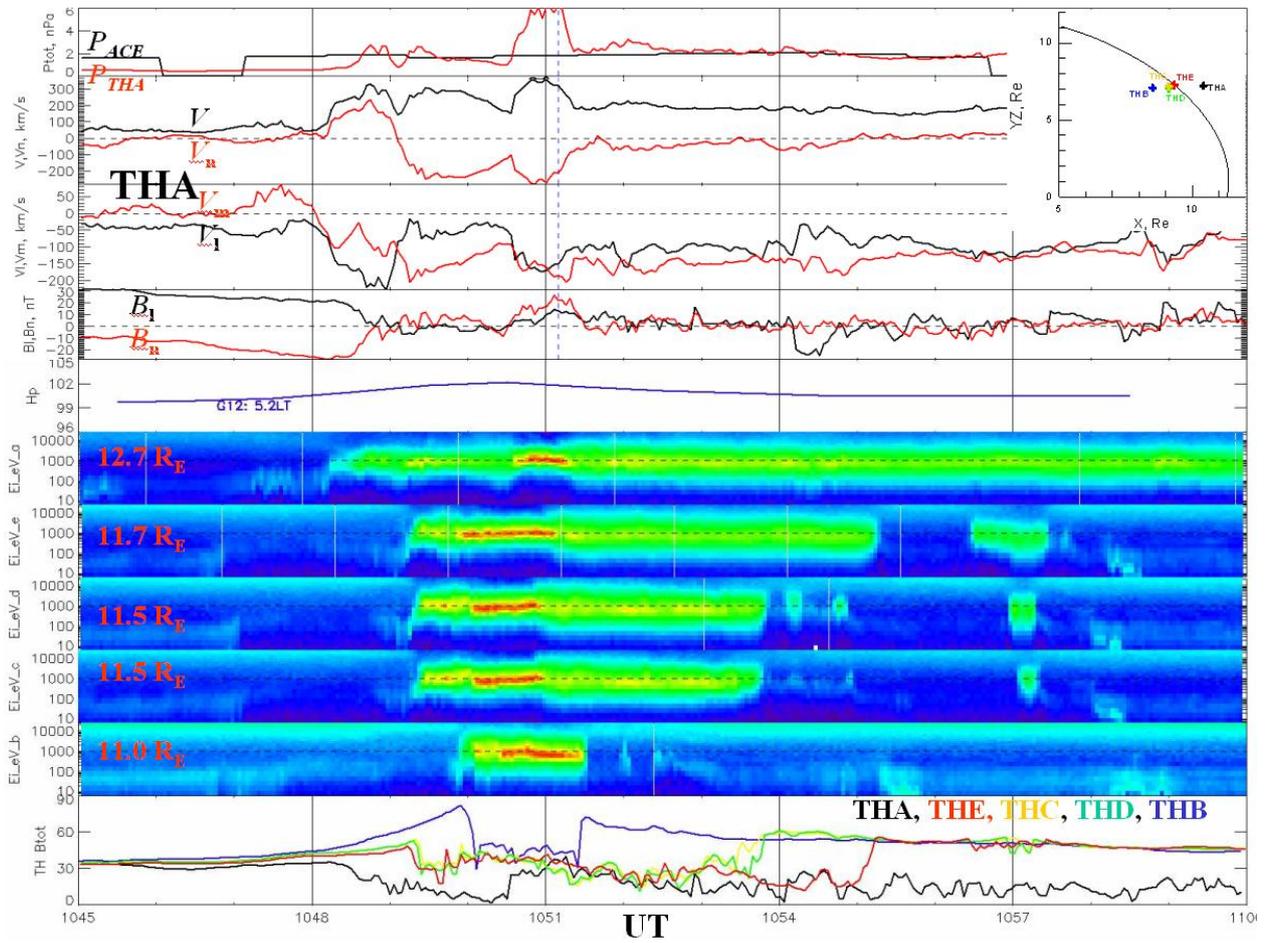

Figure 4. The same as in Figure 1 but for the THA probe on 21 July 2007. The ACE upstream data were delayed by 50 minutes. The magnetic field at geosynchronous orbit was observed by the GOES-12 satellite at 0510MLT. A fast magnetosheath plasma jet was observed by the THA probe from 1048:36 to 1052:40 UT. The magnetosheath plasma was not penetrated inside the magnetosphere as observed by the probes THB, THC and THD.

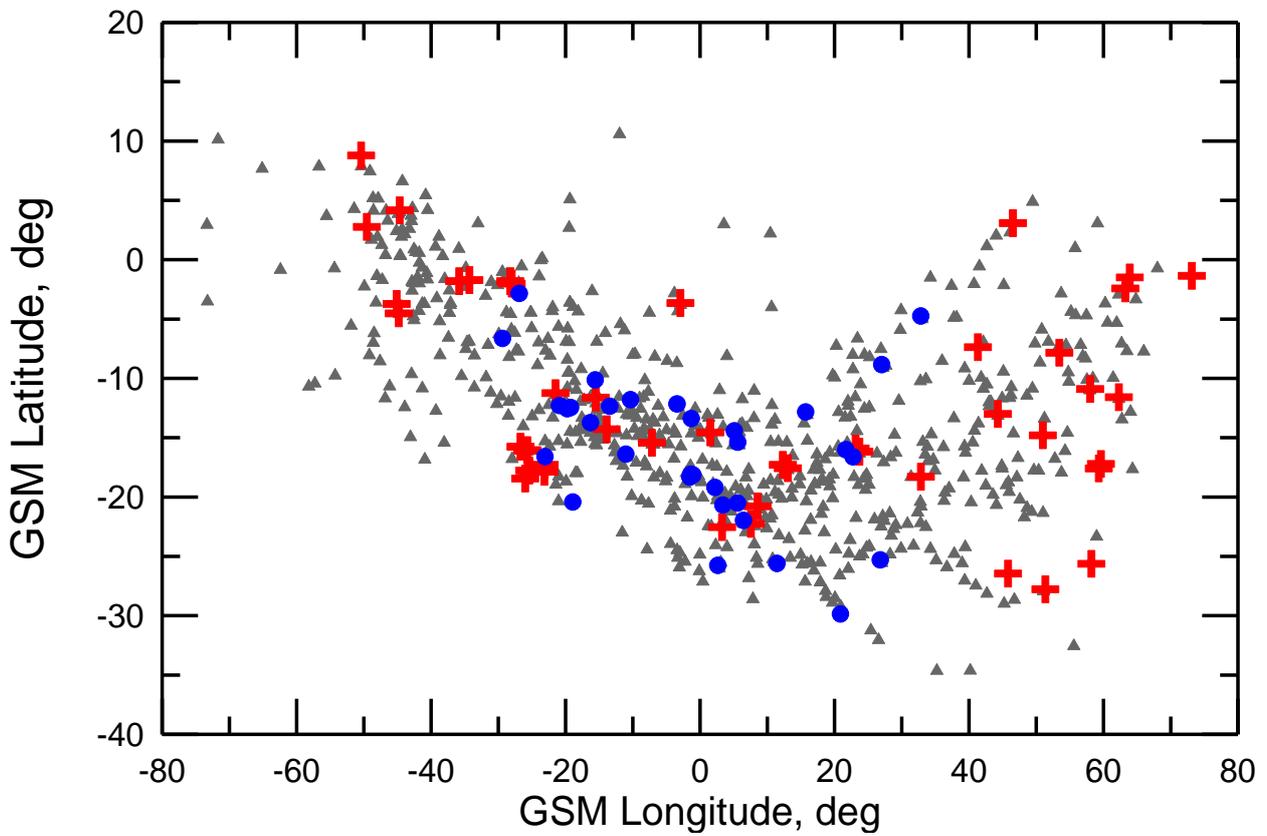

Figure 5. GSM location of large-scale magnetosheath plasma jets (gray triangles) collected from THEMIS data in 2007 – 2009. Red crosses indicate the jets, which result in plasma penetration across the magnetopause (hereafter, penetrating jets). Blue circles indicate the jets, which are not accompanied by the plasma penetration (hereafter, nonpenetrating jets). The jets are scattered quite randomly.

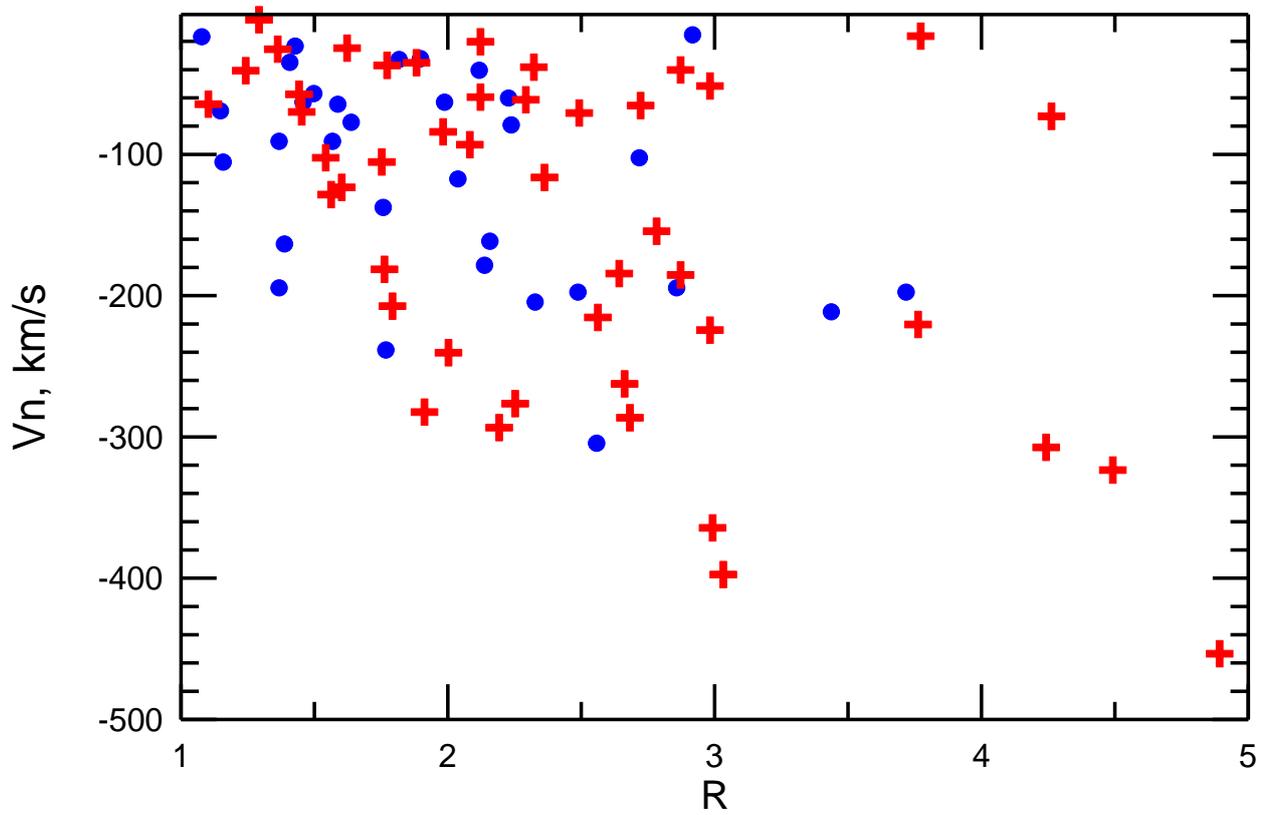

Figure 6. Scatter plot of normal component of jet velocity to the magnetopause versus the ratio *R* of total energy densities. The red crosses and blue circles indicate, respectively, penetrating and nonpenetrating jets.

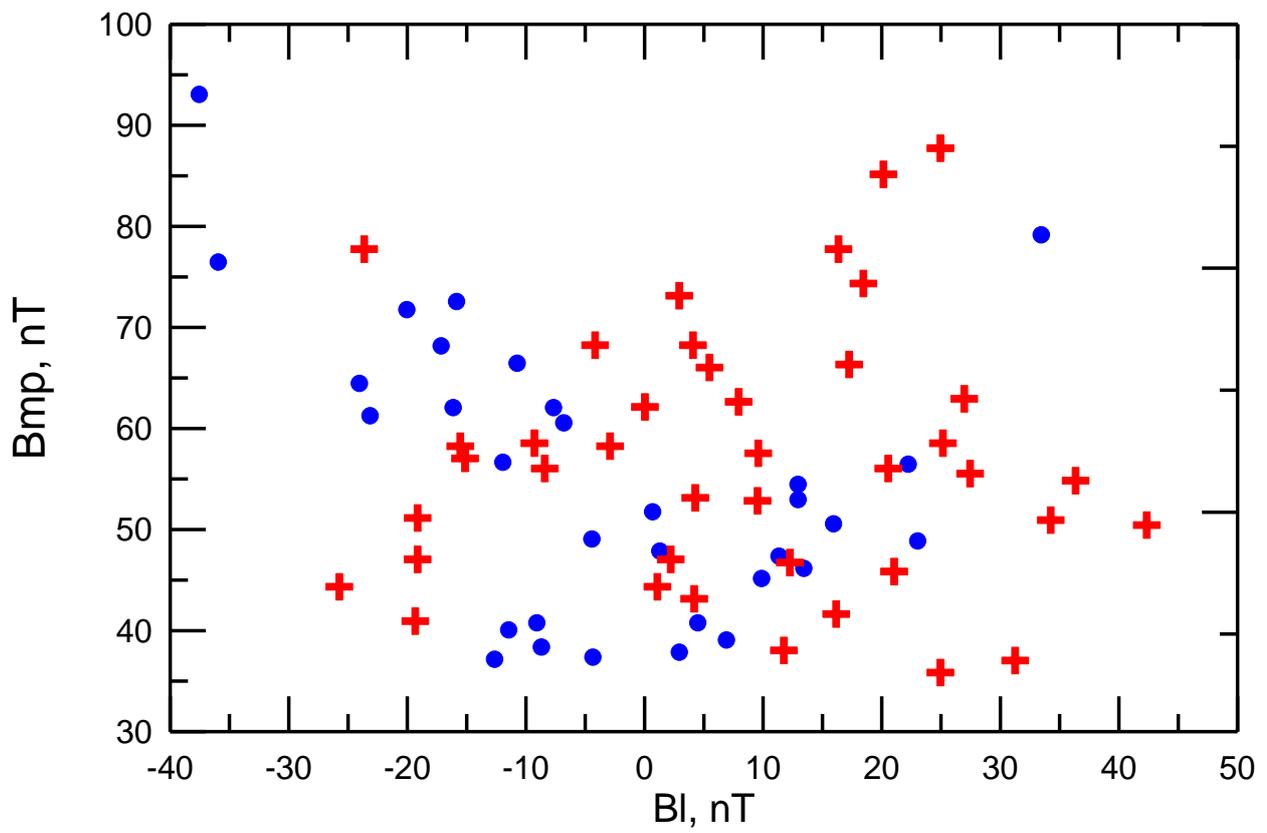

Figure 7. Scatter plot of geomagnetic field magnitude at the magnetopause (see details in the text) versus north-south component $B$l of the magnetic field in the magnetosheath plasma jet. The red crosses and blue circles indicate, respectively, penetrating and nonpenetrating jets.

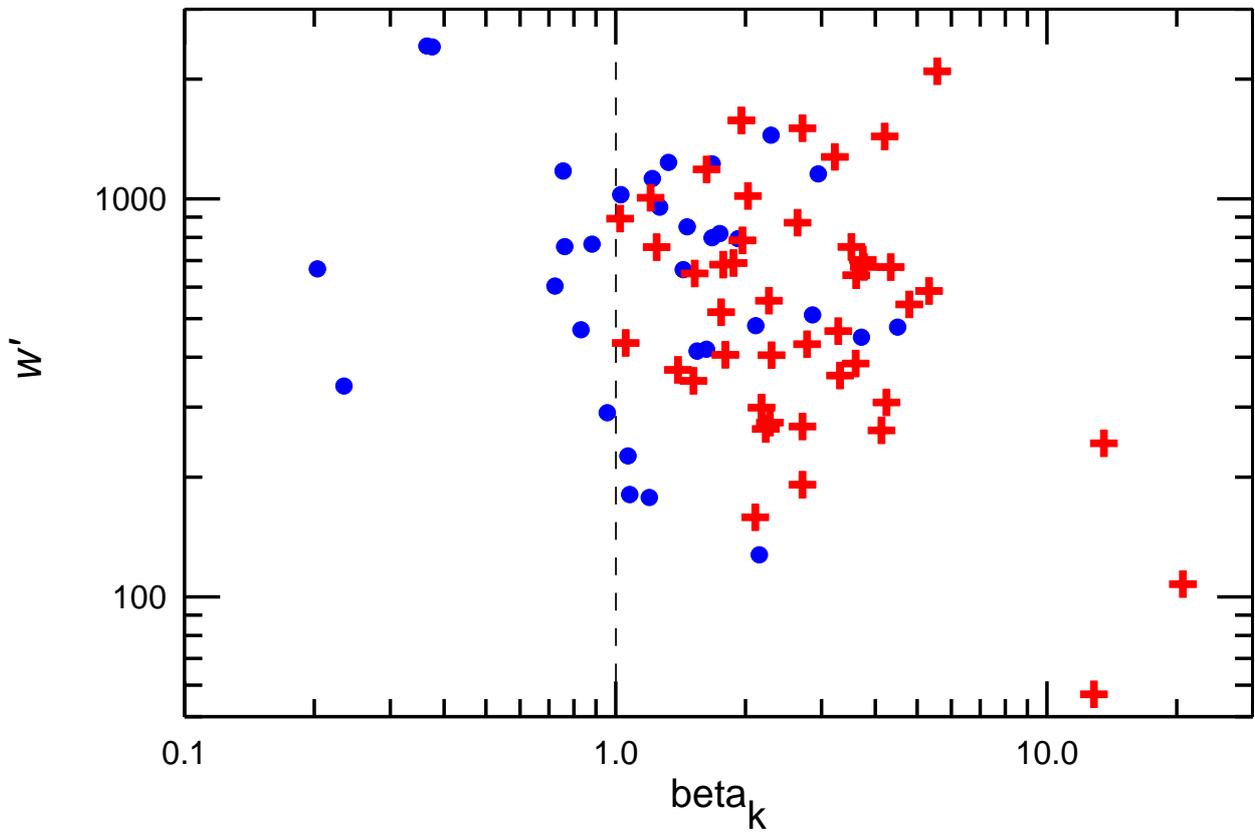

Figure 8. Scatter plot of large-scale magnetosheath plasma jets in coordinates of scale parameter *w'* (see details in the text) versus kinetic $\beta_k$. The red crosses and blue circles indicate, respectively, penetrating and nonpenetrating jets.

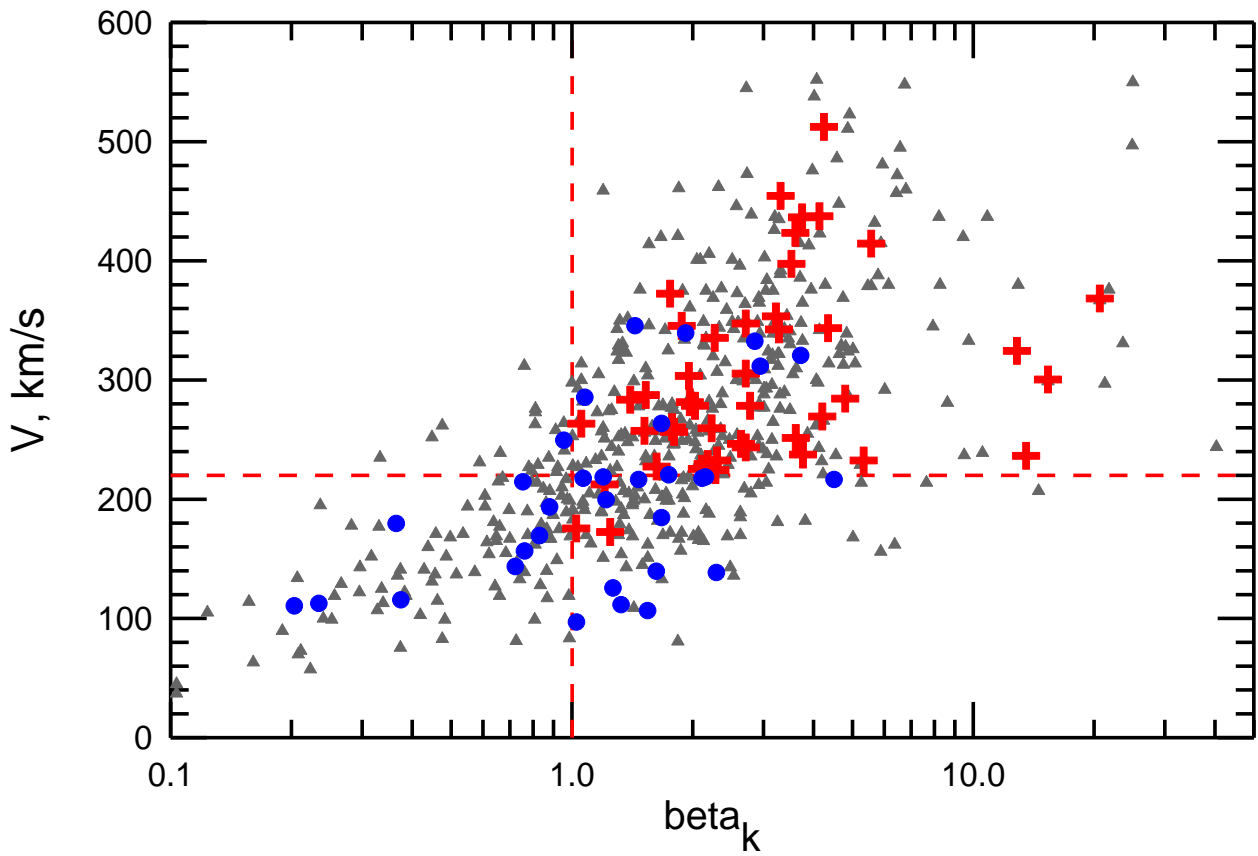

Figure 9. Scatter plot of velocity $V$ versus kinetic $\beta_k$ of the jets collected (gray triangles). The red crosses and blue circles indicate, respectively, penetrating and nonpenetrating jets. Vast majority of the penetrating jets are characterized by $\beta_k > 1$ and $V > 210$ km/s.

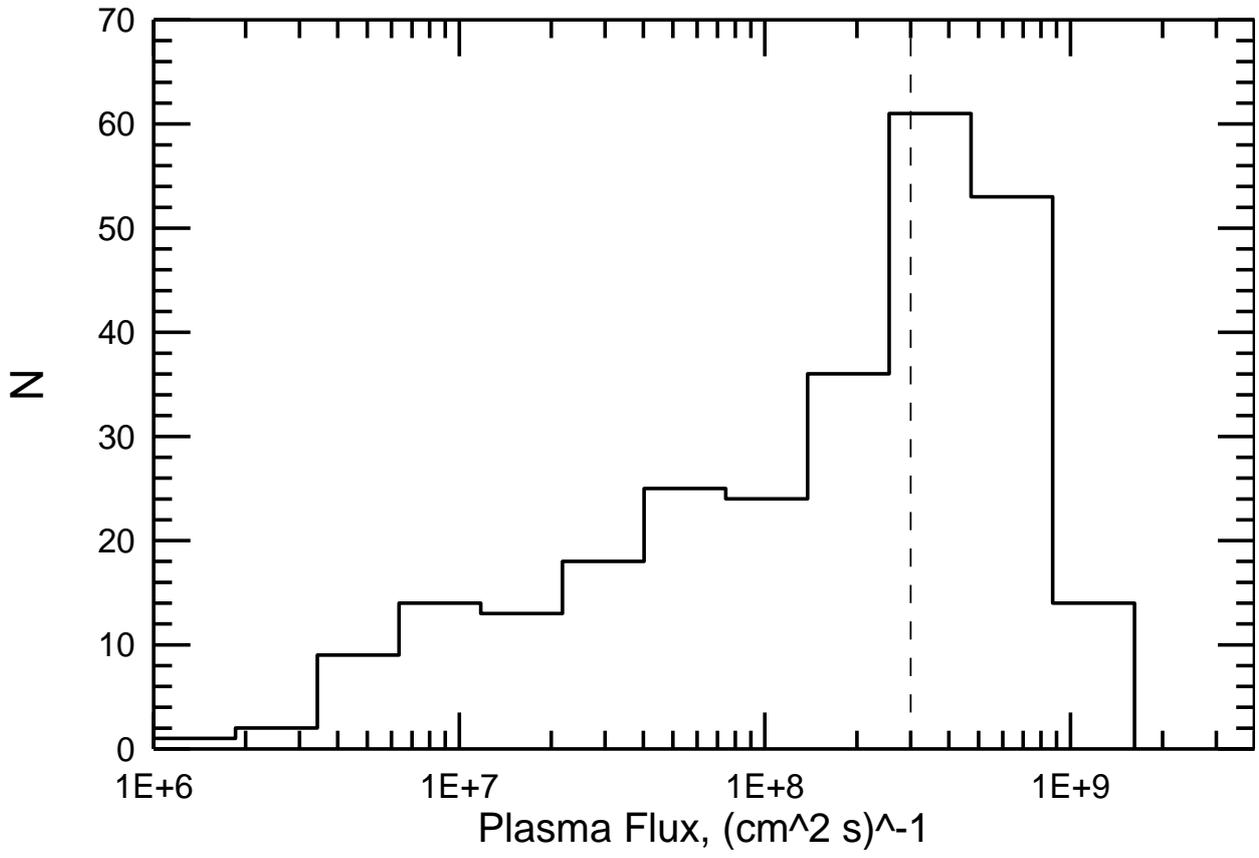

Figure 10. Statistical distribution of the plasma fluxes in the jets satisfying the penetration conditions of $\beta_k > 1$ and $V > 220$ km/s. The most probable mean plasma flux is $\sim 3*10^8$ $(cm^2\ s)^{-1}$.

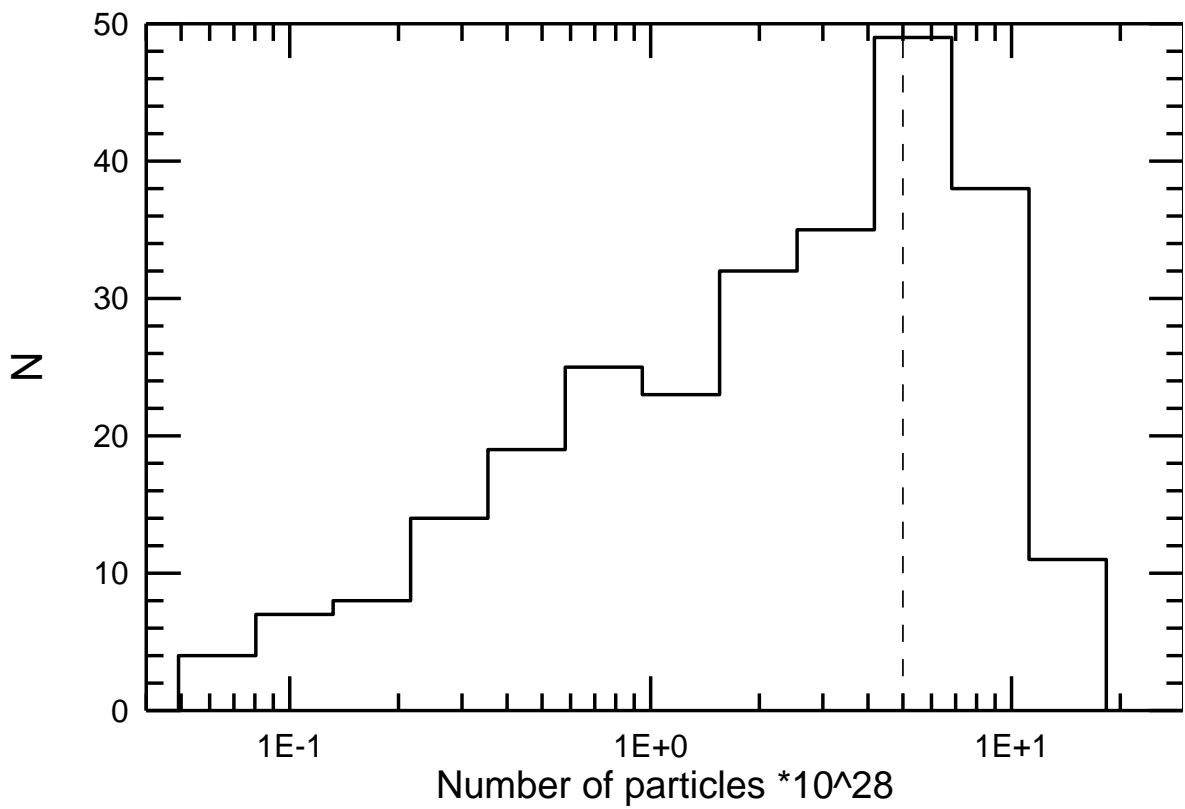

Figure 11. Statistical distribution of the number of particles carrying by penetrating jets. The average number is equal to $5*10^{28}$ particles.

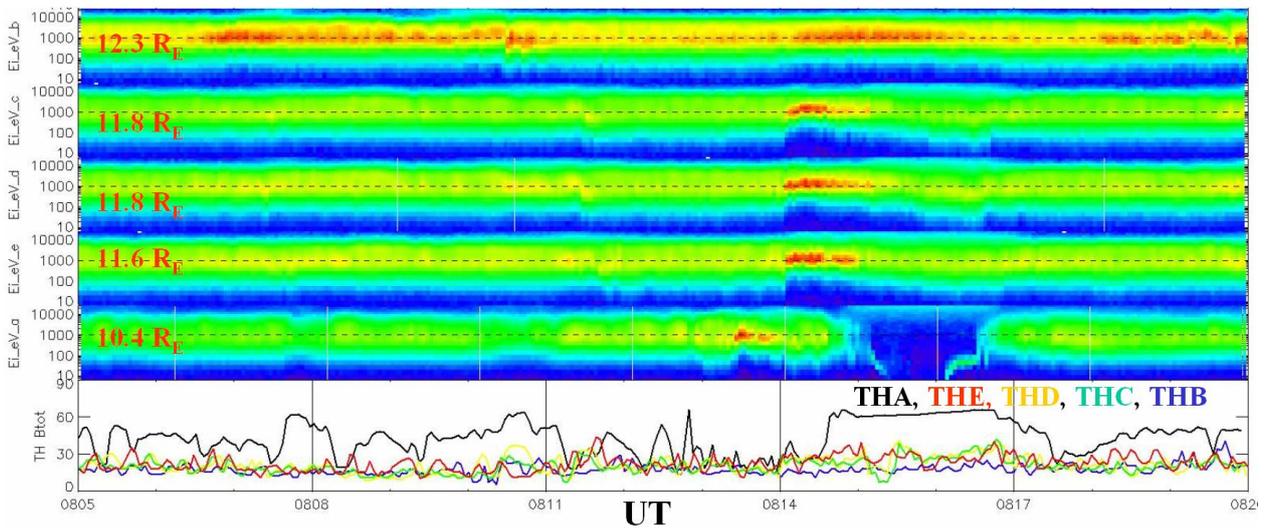

Figure 12. Observations of plasma and magnetic field by THEMIS on 23 June 2007 (from top to bottom): ion spectrograms measured by THB, THC, THD, THE and THA; magnitude of magnetic field measured by THA, THE, THD, THC and THB shown, respectively, by black, red, yellow, green and blue curves. Red numbers indicate the geocentric distance of THEMIS probes. Around 0814 UT, a fast plasma jet was observed in the magnetosheath. The key parameters of jet are presented in Table 3.